\documentstyle[12pt]{article}

\catcode`\@=11
\long\def\@makefntext#1{
\protect\noindent \hbox to 3.2pt {\hskip-.9pt
$^{{\ninerm\@thefnmark}}$\hfil}#1\hfill}              

\def\@makefnmark{\hbox to 0pt{$^{\@thefnmark}$\hss}}  

\def\ps@myheadings{\let\@mkboth\@gobbletwo
\def\@oddhead{\hbox{}
\rightmark\hfil\ninerm\thepage}
\def\@oddfoot{}\def\@evenhead{\ninerm\thepage\hfil
\leftmark\hbox{}}\def\@evenfoot{}
\def\sectionmark##1{}\def\subsectionmark##1{}}

\renewcommand{\thefootnote}{\fnsymbol{footnote}}

\newcounter{sectionc}
\newcounter{subsectionc}
\newcounter{subsubsectionc}
\renewcommand{\section}[1] {\vspace*{0.6cm}\addtocounter{sectionc}{1}
\setcounter{subsectionc}{0}\setcounter{subsubsectionc}{0}\noindent
        {\normalsize\bf\thesectionc. #1}\par\vspace*{0.4cm}}
\renewcommand{\subsection}[1] {\vspace*{0.6cm}
        \addtocounter{subsectionc}{1}
        \setcounter{subsubsectionc}{0}\noindent
        {\normalsize\it\thesectionc.\thesubsectionc. #1}
        \par\vspace*{0.4cm}}
\renewcommand{\subsubsection}[1]
{\vspace*{0.6cm}\addtocounter{subsubsectionc}{1}
        \noindent
{\normalsize\rm\thesectionc.\thesubsectionc.\thesubsubsectionc.
        #1}\par\vspace*{0.4cm}}

\newcounter{appendixc}
\newcounter{subappendixc}[appendixc]
\newcounter{subsubappendixc}[subappendixc]

\renewcommand{\appendix}[1] {\vspace*{0.6cm}
        \refstepcounter{appendixc}
        \setcounter{figure}{0}
        \setcounter{table}{0}
        \setcounter{equation}{0}
        \renewcommand{\thefigure}{\Alph{appendixc}.\arabic{figure}}
        \renewcommand{\thetable}{\Alph{appendixc}.\arabic{table}}
        \renewcommand{\theappendixc}{\Alph{appendixc}}
        \renewcommand{\theequation}
        {\Alph{appendixc}.\arabic{equation}}
        \noindent{\bf Appendix \theappendixc #1}\par\vspace*{0.4cm}}

\renewenvironment{thebibliography}[1]
        {\begin{list}{\arabic{enumi}.}
        {\usecounter{enumi}\setlength{\parsep}{0pt}
\setlength{\leftmargin 1.25cm}{\rightmargin 0pt}
         \setlength{\itemsep}{0pt} \settowidth
        {\labelwidth}{#1.}\sloppy}}{\end{list}}

\newcounter{itemlistc}
\newcounter{romanlistc}
\newcounter{alphlistc}
\newcounter{arabiclistc}

\newcommand{\fcaption}[1]{
        \refstepcounter{figure}
        \setbox\@tempboxa = \hbox{\footnotesize Fig.~\thefigure. #1}
        \ifdim \wd\@tempboxa > 6in
           {\begin{center}
        \parbox{6in}{\footnotesize
        \baselineskip=12pt Fig.~\thefigure. #1}
            \end{center}}
        \else
             {\begin{center}
             {\footnotesize Fig.~\thefigure. #1}
              \end{center}}
        \fi}

\newcommand{\tcaption}[1]{
        \refstepcounter{table}
        \setbox\@tempboxa = \hbox{\footnotesize Table~\thetable. #1}
        \ifdim \wd\@tempboxa > 6in
           {\begin{center}
        \parbox{6in}
        {\footnotesize\baselineskip=12pt Table~\thetable. #1}
            \end{center}}
        \else
             {\begin{center}
             {\footnotesize Table~\thetable. #1}
              \end{center}}
        \fi}

\def\@citex[#1]#2{\if@filesw\immediate\write\@auxout
        {\string\citation{#2}}\fi
\def\@citea{}\@cite{\@for\@citeb:=#2\do
        {\@citea\def\@citea{,}\@ifundefined
        {b@\@citeb}{{\bf ?}\@warning
        {Citation `\@citeb' on page \thepage \space undefined}}
        {\csname b@\@citeb\endcsname}}}{#1}}

\newif\if@cghi
\def\cite{\@cghitrue\@ifnextchar [{\@tempswatrue
        \@citex}{\@tempswafalse\@citex[]}}
\def\citelow{\@cghifalse\@ifnextchar [{\@tempswatrue
        \@citex}{\@tempswafalse\@citex[]}}
\def\@cite#1#2{{$\null^{#1}$\if@tempswa\typeout
        {IJCGA warning: optional citation argument
        ignored: `#2'} \fi}}

 1
 1
 1

\font\ninerm=cmr9

\input epsf
\global\arraycolsep=2pt

\begin{document}

\begin{titlepage}

\begin{flushright}
LPTENS-96/07\\
CERN-TH/96-17\\
hep-th/9601153
\end{flushright}

\vspace{1.5cm}

\begin{center}
\Large\bf Advances in Large $N$ Group Theory and the Solution of
Two-Dimensional $R^2$ Gravity
\end{center}

\vspace{0.75cm}

\begin{center}
Vladimir A.~Kazakov$^{(1)}$\\
Matthias Staudacher$^{(1),(2),}$\footnote{
This work is supported by funds provided by the European Community,
Human Capital and Mobility  Programme.}\\
{\it and}\\
Thomas Wynter$^{(1),*}$
\end{center}

\vspace{0.25cm}

\begin{center}
(1) Laboratoire de Physique Th\'eorique de\\
l'\'Ecole Normale Sup\'erieure\footnote{
Unit\'e Propre du
CNRS,
associ\'ee \`a l'\'Ecole Normale Sup\'erieure et \`a
l'Universit\'e de Paris-Sud.}\\
(2) CERN, Theory Division

\end{center}

\vspace{1.0cm}

\begin{abstract}

We review the recent exact solution of a matrix model which
interpolates between flat and random lattices.
The importance of the results is twofold:
Firstly, we have developed a new large $N$ technique capable of
treating a class of matrix models previously thought to be
unsolvable. Secondly, we are able to make a first precise
statement about two-dimensional $R^2$ gravity.
These notes are based on a lecture given at the Cargese summer
school 1995. They contain some previously unpublished results.
\end{abstract}

\vspace{1.5cm}

\centerline{\it To appear in the Proceedings of the
1995 Cargese Summer School}
\centerline{\it ``Low-dimensional Applications of Quantum
Field Theory''}

\vspace{2.0cm}

\noindent
CERN-TH/96-17\\
January 1996

\end{titlepage}

\normalsize\baselineskip=15pt
\setcounter{footnote}{0}
\renewcommand{\thefootnote}{\alph{footnote}}

\def\barint{-\hskip -11pt\int}
\def\cut{\hskip 2pt{\cal /}\hskip -8.1pt}

The large $N$ expansion\index{Large $N$ expansion}
of matrix-valued field theories
was invented more than twenty years ago by `t Hooft as
a way to treat four-dimensional
QCD\index{Large $N$ Expansion! and QCD}
with gauge group
$SU(N)$. At $N= \infty$ planar
diagrams dominate and it was hoped that this fact would
lead either to analytic results or to a reformulation
of QCD as a
string theory.\index{Large $N$ Expansion! and string theory}
While the original ideas remain
attractive, the program has not yet been successful.
It was slowly understood that $N=\infty$ field theories retain
much of the complexity of the generic $N$ case.

Some of this complexity remains even in zero and one-dimensional
matrix ``field theories''.
In a famous paper\cite{BIPZ}
by Br\'ezin, Itzykson, Parisi and Zuber
it was demonstrated that
these so-called matrix models\index{Matrix models} are
non-trivial -- but still solvable -- systems. E.g.~the zero
dimensional model\index{Matrix Models! Hermitian}
\begin{equation}
Z=\int~{\cal D}M~e^{-N\ {\rm Tr}~[ {1\over 2} M^2~ -
{}~\Sigma_{q=1}^{\infty}~t_q M^q ]},
\label{matrix}
\end{equation}
where $M$ is a $N \times N$ Hermitian matrix and the $t_q$'s
are coupling constants parametrizing a general potential,
is solved\cite{BIPZ} by changing variables to the eigenvalues
of the matrix $M$ and thereby reducing the number of degrees
of freedom\index{Large $N$ expansion|see {Matrix models}}
from $N^2$ to $N$. The latter proves possible
due to the invariance of the above action and measure
under the group $U(N)$. The model's free energy
describes at $N=\infty$
a sum over all planar diagrams $G$ of spherical topology, where
vertices $v_q$  of order $q$ (there are $\#v_q$ of them in G)
are weighted with a factor $t_q$:
\begin{equation}
\log Z~\sim~{\cal Z}=\sum_{G}~\prod_{v_q \in G}~t_q^{\# v_q}
\label{graphs}
\end{equation}
The rather non-trivial combinatorial sum (\ref{graphs})
is elegantly calculated by solving (\ref{matrix}).
At the time\cite{BIPZ} the model was considered as
a kind of ``toy'', zero-dimensional QCD (retaining
the diagrammatic structure but nothing else) to test
a new technique. A few years later, however, it became
clear that the results could be used to obtain the solution
of a rather interesting physical problem:
two-dimensional quantum gravity\index{Quantum gravity! $2D$}
\cite{DAVID,VOL,FRO}.
Indeed, it was argued that by tuning the couplings $t_q$ in
an appropriate way a continuum limit could be reached
at which the planar graphs condense to give a continuum
path integral over two-dimensional metrics $g_{ab}$ of spherical
topology:
\begin{equation}
{\cal Z}_{\rm cont}
= \int {\cal D}g_{ab}~e^{-\int d^2z \sqrt{\det g}
{}~(\mu + {1 \over \alpha} R_g ) } .
\label{contone}
\end{equation}
The distance from the critical point in the space of the $t_q$'s
turns into a continuum cosmological constant $\mu$ controlling
the area of the surfaces. We also wrote the Einstein term,
which is however known to be a constant in two dimensions.
This approach was subsequently worked out and justified
by a large number of researchers.

Thus already the simplest matrix model (\ref{matrix}) contains
non-trivial physics. Various generalisations were solved and shown
to give new physical information; e.g.~certain simple
multi matrix models describe the coupling of $c\leq 1$
conformal matter to 2D gravity (\ref{contone}). Such
generalisations usually required the development of
new techniques in order to succeed. Turning this around,
each technical advance in large $N$ theory usually allows
to address a previously inaccessible physical problem.
Keeping in mind the final goals -- like
large $N$ QCD\index{Large $N$ Expansion! and QCD}
or
string theories\index{Large $N$ Expansion! and string theory}
in physical dimensions -- a valuable strategy is, then, to
continue to enlarge our tools and methods.

An encouraging example for this strategy was given recently in
a series of papers\cite{KSWI,KSWII,KSWIII}.
We studied a generalisation of (\ref{matrix}) consisting
in the inclusion of an external
matrix field\index{Matrix Models! Hermitian in external field}
$A$ in the potential:
\begin{equation}
Z=\int~{\cal D}M~e^{-N\ {\rm Tr}~[ {1\over 2} M^2~ -
{}~\Sigma_{q=1}^{\infty}~t_q (M A)^q ]}.
\label{modmatrix}
\end{equation}
On a technical level, this model seemed for a long time
unsolvable: The external field destroys the $U(N)$ invariance
of the action. Thus none of the usual methods, whose essence
consists in reducing the number of degrees of freedom
from $N^2$ to $N$, appears applicable to (\ref{modmatrix}).
We succeeded in nevertheless finding a reduction of the
number of degrees of freedom
by developing the new method of large $N$
character expansions
\index{Character expansions| see {Large $N$ expansion}}.
This allows us to address new
physical questions. Indeed, it is easy to prove that
the perturbative expansion
in planar graphs of (\ref{modmatrix}) is given by
\begin{equation}
\log Z~\sim~{\cal Z}=
\sum_{G} \prod_{v^*_q,v_q \in G}
{t_q^*}^{\# v^*_q}\ {t_q}^{ \#v_q},
\label{DWG}
\end{equation}
where
\begin{equation}
t_q^*={1 \over q}~{1 \over N}~{\rm Tr}~A^q.
\label{adef}
\end{equation}
The sum is, as in eq.(\ref{graphs}), taken over all planar graphs
$G$ of spherical topology. But now we have an extra set of
coupling constants $t_q^*$ at our disposal: They assign weights
to the vertices $v_q^*$ of the {\it dual} lattice (there are
$\#v_q^*$ of them in $G$). Thus the $t_q$ and $t_q^*$
control the coordination numbers of the vertices and the faces
of $G$, respectively. In particular, if we set
$t_q=t_q^*=\delta_{q,4}$ the only surviving graphs in the ensemble
$\{G\}$ are regular square lattices\footnote{Of course there is no
square lattice of spherical topology; one needs to also add some
positive curvature defects to be able to close the graph into
a sphere; see below.}.
Thus, the model (\ref{modmatrix}),(\ref{DWG}) is capable of
interpolating between fluctuating random lattices and flat,
regular lattices!
In the continuum formulation, this property would be achieved by
adding higher curvature counterterms\index{Quantum gravity! $R^2$}
to the action of (\ref{contone}):
\begin{equation}
{\cal Z}_{\rm cont}
= \int {\cal D}g_{ab}~e^{-\int d^2z \sqrt{\det g}
{}~(\mu + {1 \over \alpha} R_g + {1\over \beta_0} R_g^2 + \ldots) } .
\label{conttwo}
\end{equation}
Tuning the bare coupling $\beta_0$ to zero clearly suppresses any
non-flat metric. Our model therefore furnishes a
precise invariantly regularized definition of 2D higher
curvature gravity\index{Quantum gravity! Higher curvature}.
Before discussing the physics of the latter, let us first
explain the steps that lead to a solution of the apparently
intractable matrix model (\ref{modmatrix}). We will only sketch
the derivation (and even the results!); a much more detailed
discussion
can be found in the original papers\cite{KSWI,KSWII,KSWIII}.

The idea is to expand the potential in
(\ref{modmatrix}) in terms of
the characters of the product matrix $M A$.
It is clear that this is possible since the potential is
a class function on the group (i.e.~it only depends
on the eigenvalues of the matrix $M A$). What is less
obvious (proven in our first work\cite{KSWI})
is that the expansion
coefficients can themselves be written as the characters
of an auxiliary external matrix $B$
generating the couplings $t_q$, just as $A$ generates the
couplings $t_q^*$ (see eq.(\ref{adef})):
\begin{equation}
t_q={1 \over q}~{1 \over N}~{\rm Tr}~B^q.
\label{bdef}
\end{equation}
One then has the expansion
\begin{equation}
e^{N {\rm Tr}~\Sigma_{q=1}^{\infty}  t_q (MA)^q}=
c\,\sum_R\chi_R(B)\ \chi_R(MA),
\label{potchar}
\end{equation}
with $c$ a numerical constant. The characters are defined by the Weyl
formula
\begin{equation}
\chi_{\{h\}}(B)={{\rm det}_{_{\hskip-2pt (k,l)}}(b_k^{h_l})\over
\Delta(b)},
\label{eigchar}
\end{equation}
where the $b_i$ are the eigenvalues of the matrix $B$,
$\Delta(b)$ is the Vandermonde determinant
of the eigenvalues,
the set of $\{h\}$ are a set of ordered, increasing,
non-negative integers, and the
sum over $R$ is the sum over all such sets.  The $R$'s label
representations of the group $U(N)$ and the sets of integers $\{h\}$
are the usual Young tableau weights defined by
$h_i=i-1+\#$boxes in row $i$ (the index $i$ labels the rows in
the Young tableau, $i=1$ corresponding to the lowest row). Note
that the restriction on the allowed
Young tableaux
\index{Young tableau|see {Large $N$ expansion}} that any row must
have at least as many boxes as the row below implies that the
$\{h_i\}$ are a set of increasing integers:
$h_{i+1}>h_i$. Substituting
equation (\ref{potchar}) into the integral in equation
({\ref{modmatrix}),
we can now do the angular integration using the key identity
\begin{equation}
\int\,({\cal
D}\Omega)_H~\chi_R(\Omega M\Omega^{\dagger}A)=
d_R^{-1}~\chi_R(M)~\chi_R(A),
\label{angular}
\end{equation}
where $d_R$ is the dimension of the representation given,
up to a constant, by
$d_R \sim \Delta(h)$), and arrive, after performing
a Gaussian integral over the eigenvalue degrees of freedom,
at the expression
\begin{equation}
Z=c\,\sum_{\{h^e,h^o\}}
{\prod_i(h^e_i-1)!!h^o_i!!\over
\prod_{i,j}(h^e_i-h^o_j)}~\chi_{\{h\}}(A)~\chi_{\{h\}}(B),
\label{IzDiFr}
\end{equation}
where $c$ is an immaterial constant.
The sum is taken over a subclass of so-called even representations.
These are defined as possessing an equal number of even weights
$h^{e}_{i}$ and odd weights $h^{o}_{i}$ (since the mentioned Gaussian
integration vanishes if the latter condition is not satisfied).
The formula (\ref{IzDiFr}) was originally discovered by Itzykson and
Di
Francesco\cite{IDiF}
by summing up ``fatgraphs'', using purely combinatoric
and group theoretic arguments. We observe that the matrix model
(\ref{modmatrix}) is thus reformulated as a sort of
``statistical mechanics model'' in Young weight space.
The important fact is that there are only $N$ weights $h_i$;
therefore
the reduction to $N$ degrees of freedom is achieved.

The expansion (\ref{IzDiFr}) can be further generalised.
Consider the matrix model\index{Matrix models! complex  in external
fields}
\begin{equation}
Z=\int~{\cal D}\phi~e^{-N~{\rm Tr}~[{1 \over 2} \phi \phi^{+}~
-~\sum_{k=1}^{\infty}~g_k
(\phi A \phi^{+} B)^k]},
\label{cmatrix}
\end{equation}
where $\phi$ is a {\it complex} $N\times N$ matrix.
Introducing a third external matrix field $C$ and defining,
in analogy with (\ref{adef}),(\ref{bdef}),
\begin{equation}
g_k={1 \over k}{1 \over N}~{\rm Tr}~C^k ,
\label{cdef}
\end{equation}
one finds by the method outlined above the character expansion
\begin{equation}
Z=c\,\sum_{\{ h \}}
{\prod_i h_i! \over
\Delta(h) }~\chi_{\{h\}}(A)~\chi_{\{h\}}(B)~\chi_{\{h\}}(C) ,
\label{cargese}
\end{equation}
where this time the sum extends over {\it all} representations.
In the special case $g_k = \delta_{k,2}$ the earlier expansion
(\ref{IzDiFr}) is recovered\footnote{For this special case the
correspondence between the hermitian model (\ref{modmatrix})
and the complex model (\ref{cmatrix}) was already noted
previously\cite{IDiF}.
For a graphical explanation of this correspondence
see our second work\cite{KSWII}.}.

The reduction of the number of degrees of freedom is
only a {\it conditio sine qua non}. One next has to take
the large $N$ limit of the expansion. The basic idea is the
same as for the original model (\ref{matrix})
(see\cite{BIPZ}), with Young
weights replacing eigenvalues:
The weights ${1 \over N} h_i$ are assumed to freeze into
a smooth, stationary distribution $dh~\rho(h)$, where $\rho(h)$
is a probability density normalized to one.
The details, however, turn out to be much more involved.
Some simple first examples were worked out in our
first paper\cite{KSWI}.
An unpleasant feature is that, while the saddlepoint
always exists, the support of the density $\rho(h)$
does not necessarily remain on the real axis for completely
arbitrary couplings $t_q$,$t_q^*$,
complicating the general analysis in a significant way.
However, if we restrict
our attention to models in which the matrices $A$ and $B$
are such that traces of all odd powers of $A$ and $B$ are zero
the problem does not arise: $t_{2q+1}=t_{2q+1}^*=0$. This
means that our random surfaces are made from vertices and faces
with even coordination numbers.
Thus it is easiest to consider
surfaces made up from squares (as opposed to, say, triangles):
$t_q^* = \delta_{q,4}$.
A weight $t_{2 q}={1 \over 2 q} {1 \over N}
{\rm Tr} B^{2 q}$ is assigned
whenever $2 q$ squares meet at a vertex (see Fig.~1).

\vskip 20pt
\epsfbox{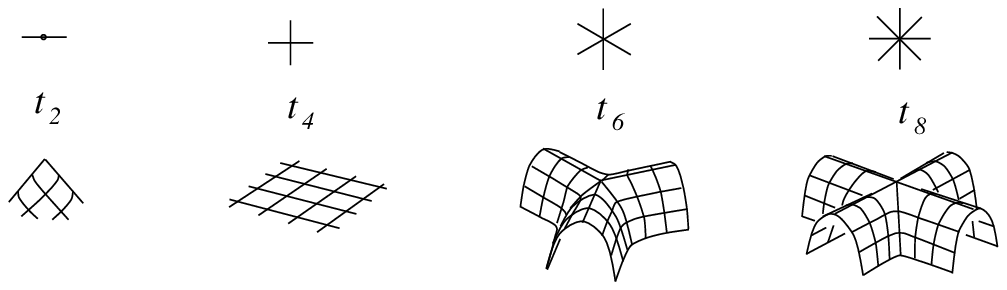}
\centerline{{\bf Fig.~1} Flat space and curvature defects.}
\vskip 30pt

This model is clearly capable of describing
the transition from flat to random graphs.
The sum (\ref{DWG}) over spherical lattices $G_4$
built from square plaquettes becomes
\begin{equation}
{\cal Z}=
\sum_{G_4} \prod_{v_{2q} \in G}  {t_{2q}}^{ \# v_{2q}},
\label{DWGG}
\end{equation}
where $v_{2q}$ are the vertices where $2q$ plaquettes meet
and $\# v_{2q}$ are the numbers of such
vertices in the given graph $G_4$.
In order to investigate (\ref{IzDiFr}) in the large $N$ limit,
one attempts to locate the stationary point.
This leads to the saddlepoint equation
\begin{equation}
2F(h)+\barint_0^a\ dh'\ {\rho(h') \over h-h'}= -\ln h.
\label{sdpt}
\end{equation}
Here $F(h)$ denotes the large $N$ limit
$F(h_k) \rightarrow F(h)$ of
the variation of the character $\chi_{\{h\}}(B)$
in eq.(\ref{IzDiFr}):
\begin{equation}
F(h_k)=2{\partial \over \partial h^e_k}~\ln\ {\chi_{\{{h^e\over
2}\}}({\bar b})
\over \Delta(h^e)}.
\label{defF}
\end{equation}
where we have also used that the
matrix $B$ will satisfy ${\rm Tr} B^{2q+1}=0$ if we
introduce a ${N\over 2}\times{N\over 2}$ matrix
${\bar b}^{1 \over 2}$ in terms
of which $B$ and the character $\chi_{\{h\}}(B)$ are given by
\begin{equation}
B=\left[\matrix{{\bar b}^{1 \over 2}&0\cr
0&-{\bar b}^{1 \over 2} \cr}\right]\quad {\rm and}
\quad
\chi_{\{h\}}(B)=
    \chi_{\{{h^e\over 2}\}}({\bar b})\chi_{\{{h^o-1\over 2}\}}({\bar
b})
    \,\,{\rm sgn} \bigl[\prod_{i,j}(h^e_i-h^o_j)\bigr].
\label{Bb}
\end{equation}
The variation of the character of the matrix $A$ is easily
computed directly because of the simple choice
$t_q^*=\delta_{q,4}$.
As has been discussed in detail before\cite{KSWI},
the saddlepoint equation (\ref{sdpt}) actually does
not hold on the entire interval $[0,a]$, but only on an interval
$[b,a]$ with $0 \leq b \leq 1 \leq a$:
Assuming the equation to hold on $[0,a]$ would violate the
implicit
constraint $\rho(h) \leq 1$ following from the restriction
$h_{i+1}>h_i$.
The density is in fact exactly saturated at its maximum value
$\rho(h)=1$
on the interval $[0,a]$.
It is useful to introduce in addition the
weight resolvent
$H(h)$ as follows:
\begin{equation}
H(h)=\int_0^a dh'\ {\rho(h') \over h-h'}.
\label{res}
\end{equation}
We found the weight resolvent $H(h)$ to be very closely
related to the
standard matrix model {\it eigenvalue} resolvent
(see\cite{BIPZ}). It provides
a direct link between the statistical distribution
of Young weights and the correlators of the model:
\begin{equation}
{1\over N}~{\rm Tr} M^{2q}=
{1 \over q}\oint\,{dh\over 2\pi i}~h^q~e^{qH(h)}.
\label{trMtwoq}
\end{equation}
Here the contour encircles the cut of $e^{H(h)}$.

The solution of (\ref{sdpt}) evidently requires knowing
the function $F(h)$.
A rather general method for its determination has been
one of the main technical achievements of our work.
The method of functionally determining an $N=\infty$
character might prove very useful for other applications.
We found the following simple result:
Further introduce the function $G(h)$ as
\begin{equation}
G(h) = e^{H(h) +F(h)},
\label{defG}
\end{equation}
where, again, the contour encircles the cut of $e^{H(h)}$.
Its importance stems from the fact that it
relates\cite{KSWI} in
a simple way the introduced functions
$F(h)$,$H(h)$ and the coupling constants $t_{2 q}$ :
\begin{equation}
t_{2q}={1\over q}\oint\,{dh\over 2\pi i}~G(h)^q.
\label{tqG}
\end{equation}
It is then easy to deduce, by changing variables
from $h$ to $G$, the expansion
\begin{equation}
h-1 = \sum_{q=1}^Q{t_{2 q}\over G^q} +
\sum_{q=1}^{\infty}
a_q~G^q.
\label{hGexp}
\end{equation}
By considering\cite{KSWII} the alternative representation of the
Weyl character as a determinant of Schur polynomials
one derives\footnote{One also finds the relation
$a_q=
{2q\over N}{\partial\over \partial t_{2q}}\ln\Bigl(
\chi_{\{{h^e\over 2}\}}({\bar b})\Bigr)$.
As mentioned before\cite{KSWIII}, rewriting eq.(\ref{hGexp})
with this expression for $a_q$ suggests a possible
relationship to integrable hierarchies of differential
equations. This conjecture was developed in several
conversations with I.~Kostov.} that
the coefficients $a_q$ of the positive powers of $G$ in
(\ref{hGexp})
are directly related to correlators of the matrix model
dual to (\ref{modmatrix}), i.e.~the model with
$t_q \leftrightarrow t_q^*$:
\begin{equation}
a_q=
\langle {1\over N}{\rm Tr}~(\tilde M B)^{2q} \rangle.
\label{dualcorr}
\end{equation}
We have also assumed for the moment that only a finite number $Q$
of couplings are non-zero (i.e.~$t_{2 q}=0$ for $q>Q$).
Furthermore, we were able to show in our second work\cite{KSWII}
that (\ref{hGexp}) implies the functional equation
\begin{equation}
e^{H(h)}={(-1)^{(Q-1)}h\over t_Q}\prod_{q=1}^Q G_q(h),
\label{HGprod}
\end{equation}
where the $G_q(h)$ are the first $Q$ branches of the multivalued
function
$G(h)$ defined through (\ref{hGexp}) which map the point $h=\infty$
to
$G=0$. The resulting picture of the analytic structure of
$G(h)$ is as follows: the couplings $t_{2 q}$ determine the
number of sheets attached to the physical sheet by the cut
of $e^{F(h)}$; this number is $Q$ for a finite number of non-zero
couplings. In turn, the parameters $a_q$ determine the sheets
attached to the physical sheet by the cut of $e^{H(h)}$.
This picture is easily verified (and was in fact
discovered in this way) for the rather trivial
cases where the potential in eq.(\ref{modmatrix}) is at
most quadratic. The case $Q=2$ is shown in Fig.~2.

\vskip 20pt
\hskip 30pt \epsfbox{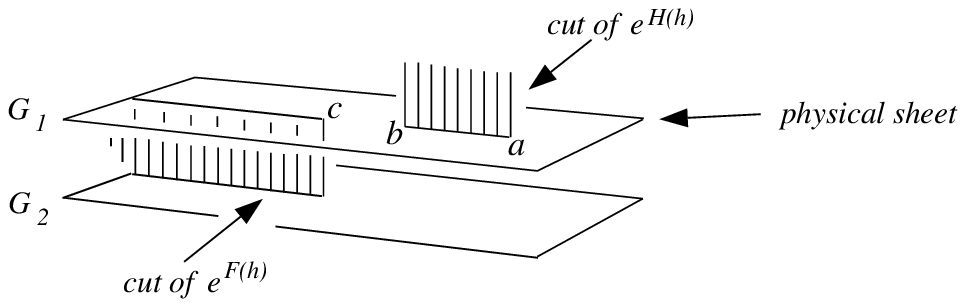}

\centerline{{\bf Fig.~2} Sheet and cut structure of $G(h)$ }
\vskip 30pt
\hskip -19pt

The saddlepoint equation (\ref{sdpt}),
together with (\ref{HGprod}),
defines a well-posed Riemann-Hilbert\index{Riemann-Hilbert problem}
problem. It was solved exactly
and
in explicit detail\cite{KSWII} for the case $Q=2$, where the
Riemann-Hilbert
problem is succinctly written in the form
\begin{eqnarray}
2F(h)+\cut H(h)=&-\ln h \nonumber \\
2\cut F(h)+H(h)=& \ln (-{t_4\over h}),
\label{RH}
\end{eqnarray}
The first equation is the saddlepoint equation
(\ref{sdpt}) with $\cut H(h)$ denoting the real part of
$H(h)$ on the cut $[b,a]$ (the righthand cut in
Fig.~2). The second equation is (\ref{HGprod}) with
$\cut F(h)$ denoting the
real part of $F(h)$ on a cut $[-\infty,c]$ with
$c<b$ (the lefthand cut in Fig.~2). This case corresponds
to an ensemble of squares being
able to meet in groups of four (i.e.~flat points with weight $t_4$)
or two (i.e.~positive curvature points with weight $t_2$)
(see Fig.~1).
We termed the resulting surfaces
``almost flat''. It turned out that all the introduced functions
could be found explicitly in terms of
elliptic functions.
E.g.~the density satisfying (\ref{sdpt}) is given by
\begin{equation}
\rho(h)={1\over K}{\rm sn}^{-1}\bigl(\sqrt{{a-h\over a-b}},k\bigr)
\quad {\rm with} \quad
k=\sqrt{{a-b\over a-c}},
\label{density}
\end{equation}
where sn$^{-1}$ is an inverse Jacobi elliptic function and
$K$ is the complete elliptic integral of the first kind
(depending on the modulus $k$). The expressions for the
cutpoints $a,b,c$ as well as the expressions for
$H(h),F(h),G(h)$ and the physical correlators $a_q$
(see (\ref{dualcorr})) were found as well.]

The resulting surfaces are very different from pure
gravity. The spherical partition function (\ref{DWGG})
consists of flat cylinders pinched shut at the two ends,
resulting in precisely four defects. Very short,
highly twisted cylinders dominate the continuum limit.
By exactly calculating the correlators $a_q$
(see eq.(\ref{dualcorr})) we also analytically
solved the
combinatorial problem of surfaces with a single
negative curvature insertion of arbitrary degree
balanced by a gas of positive defects.

To analyse the problem of the transition from flat to random
lattices, we merely need to perturb our almost flat lattices by {\it
any}
operator containing negative curvature. This physical observation
allows us to extend the $Q=2$ solution in a simple way.
Choosing the weights to be
\begin{equation}
t_2=\sqrt{\lambda}~t,~t_4=\lambda,~t_6=\lambda^{3 \over2}
{}~{\beta^2 \over t},~...~
t_{2q}=\lambda^{q \over 2}~({\beta^2 \over t})^{(q-2)},
\label{weights}
\end{equation}
the expansion (\ref{hGexp}) becomes
\begin{equation}
h-1={t_2\over G}+{t_4\over G~(G-\epsilon)}+{\rm\ positive\ powers\
of\ } G,
\label{hGfull}
\end{equation}
where $\epsilon={t_4 \over t_2} \beta^2$.
The derivation of the functional equation (\ref{HGprod}) is easily
modified (note that we now have an infinite number $Q=\infty$
of weights) to give
\begin{equation}
e^{H(h)}={h\over \epsilon t_2-t_4}~G_1(h)~G_2(h).
\label{HGgrav}
\end{equation}

The essential point is that one keeps the property that
only one other sheet $G_2(h)$ is
attached to the physical sheet $G_1(h)$
by the cut of $e^{F(h)}$ (see again Fig.~2).
The only difference is that the semi-infinite
cut $[-\infty,c]$ becomes
a finite cut $[d,c]$.
This results in modifying the Riemann-Hilbert problem
\index{Riemann-Hilbert problem} (\ref{RH})
to
\begin{eqnarray}
2F(h)+\cut H(h)=&-\ln h \nonumber\\
2\cut F(h)+H(h)=&\ln ({\epsilon t_2 - t_4 \over h }),
\label{RHgrav}
\end{eqnarray}
It may still be explicitly solved in terms of elliptic
functions; e.g.~the density (\ref{density}) is generalised
to
\begin{equation}
\rho(h)=
{u \over K} - {i \over \pi}~\ln \Bigg[
{\theta_4\big({\pi \over 2 K} ( u - i v),q \big)
\over
 \theta_4\big({\pi \over 2 K} ( u + i v),q \big)} \Bigg]
\label{moddens}
\end{equation}
with
\begin{equation}
q=e^{- \pi {K' \over K}},
\quad {\rm and} \quad
k=\sqrt{{(a-b)(c-d) \over (a-c)(b-d)}},
\label{qkdef}
\end{equation}
where $K$ and $K'$ are the complete elliptic integrals
of the first kind with respective moduli $k$
and $k'=\sqrt{1-k^2}$ and
$u$ and $v$ are given by
\begin{equation}
u={\rm sn}^{-1}\bigl(\sqrt{{(a-h)(b-d) \over (a-b)(h-d)}},k\bigr)
\quad {\rm and} \quad
v={\rm sn}^{-1}\bigl(\sqrt{{a-c \over a-d}},k'\bigr).
\label{uvdef}
\end{equation}
Again, the cutpoints $a,b,c,d$ can be explicitly obtained as
functions of the couplings $\lambda,\beta,t$.

Let us also mention that the weights (\ref{weights}) can be
further generalised while keeping the quadratic (two-sheeted)
structure of the function $G(h)$. The idea is to
shift away the simple pole at $G=0$ in the
expansion (\ref{hGfull}), i.e.~
\begin{equation}
h-1={c_1\over G - \epsilon_1}+{c_2 \over G-\epsilon_2}+
{\rm\ positive\ powers\
of\ } G.
\label{hGmod}
\end{equation}
Here $c_1$ and $c_2$ are two constants given by
$c_1={t_2 \epsilon_2 - t_4 \over \epsilon_2 - \epsilon_1}$
and
$c_2={t_4 - t_2 \epsilon_1 \over \epsilon_2 - \epsilon_1}$.
It leads to the
Riemann-Hilbert problem\index{Riemann-Hilbert problem}
\begin{eqnarray}
2F(h)+\cut H(h)=&-\ln h \nonumber\\
2\cut F(h)+H(h)=&
\ln ({(\epsilon_1 + \epsilon_2) t_2 - t_4 \over h } +
\epsilon_1 \epsilon_2 {h-1 \over h}),
\label{RHmod}
\end{eqnarray}
The weights generalizing (\ref{weights}) by one extra parameter
can be found explicitly by expanding (\ref{hGmod}) in inverse
powers of $G$. We have not worked out this further explicitly
solvable case in detail. Analysing it would furnish an
interesting universality check of our result.

Having at hand the explicit solution of the model for
the weights (\ref{weights}), we are in a position
to analyse the problem of discrete 2D $R^2$ gravity.
With these weights it is easy to prove, using Euler's theorem, that
the partition sum (\ref{DWGG}) becomes
\begin{equation}
{\cal Z}(t,\lambda,\beta) =
t^4~\sum_{G_4}~\lambda^A~\beta^{2 (\# v_{2}-4)},
\label{PART}
\end{equation}
where $A$ is the number of plaquettes of the graph $G$ and $\#v_2$
the
number of positive curvature defects. Note that the latter are
balanced by a gas of negative curvature defects, whose individual
probabilities are given in (\ref{weights}).

We expect this model to describe pure gravity in a
sufficiently large interval of $\beta$, after tuning
the bare cosmological constant $\lambda$ (controlling the number
of plaquettes) to some critical
value $\lambda_c(\beta)$.
On the other hand, for $\lambda$ fixed and $\beta=0$ we
entirely suppress curvature defects except for the four positive
defects
needed to close the regular lattice into a sphere. It is thus clear
that $\beta$ is the precise lattice analog of the bare curvature
coupling\index{Quantum gravity! $R^2$}
$\beta_0$ in the continuum path integral (\ref{conttwo}).
The phase $\beta=0$
of ``almost flat'' lattices -- very different from pure gravity --
studied in detail in the second
paper\cite{KSWII} was discussed above.

Let us now summarize the main physical results following from the
exact solution (for general $\lambda$ and $\beta$) of this model:

1. A long debated question was whether models of the present type
undergo a ``flattening'' phase transition
\index{Phase transition|see {Quantum gravity}}
at a finite, non-zero
critical value of $\beta=\beta_c$.
The weak coupling region $\beta > \beta_c$
would then correspond to the standard phase of pure gravity while
a putative novel ``smooth'' phase of gravity might exist either
at $\beta = \beta_c$ or in the entire interval
$0 \leq \beta \leq \beta_c$.
This would constitute an existence proof of {\it continuum}
2D $R^2$ gravity.
We found analysing the exact solution, to the contrary, that
{\it there is no ``flattening'' phase transition at non-zero
$\beta$}. For any given $\beta$ we find the powerlike
scaling of standard pure gravity on large scales.
This means that no matter how flat the system is on small scales (of
the order
of $\beta^{-{1 \over 2}}$), it destabilizes in the infrared
into the familiar ensemble of highly fractal ``baby-universes''.

2. The dependence of the partition sum (\ref{PART})
on $\beta$ and the
lattice cosmological constant $\lambda$ in the vicinity
of the flat phase $\beta \sim 0$ and close to
$\lambda \sim \lambda_c$ is given by a simple,
(presumably) universal scaling function $f(x)$
(defined through ${\cal Z}(t,\lambda,\beta)
= {4t^4\over 15\beta^2}~f(x)$)
reflecting the transition
from flat space to pure gravity:
\begin{equation}
f(x)=x^6 - {5 \over 2}~x^4 + {15 \over 8}~x^2 - {5 \over 16} -
     x~\big(x^2-1\big)^{5 \over 2},
\label{scalf}
\end{equation}
where the scaling variable $x$ is given, to leading order, by
\begin{equation}
x={\sqrt{2} \over \pi}~{1-\lambda \over \beta}.
\label{scalx}
\end{equation}
We can distinguish the following features:

(a) There is a degree ${5 \over 2}$ singularity at $x=1$, correctly
reproducing the universal string susceptibility exponent
$\gamma_s = -{1 \over 2}$ of pure gravity\cite{DAVID,VOL}.
In view of eq.(\ref{scalx}),
the critical value of the lattice cosmological
constant $\lambda$ is therefore given to leading order by
$\lambda_c=1-{\pi \over \sqrt{2}} \beta + {\cal O}(\beta^2)$.
Therefore (see (\ref{PART})),
the characteristic growth of the random surfaces as
a function of the lattice area $A$ (= number of plaquettes)
is given by
\begin{equation}
{\cal Z}(t,A,\beta)~\sim~{t^4\over\beta^{9\over 2}}~e^{{\pi
\over \sqrt{2}}~\beta~A}~A^{-{7 \over 2}}.
\label{growth}
\end{equation}
For any non-zero $\beta$ we do have exponential growth of the
number of surfaces, but one has to go to larger and larger scales
(i.e.~use more and more plaquettes) to be able to take the continuum
limit.
If $\beta$ is exactly zero there is no longer any exponential growth
and
no pure gravity continuum limit is possible. The prefactor
$\beta^{-{ 9 \over 2}}$ is found in the exact calculation in
section 5; we are not sure whether it is universal.

(b) We further see that taking $\beta \rightarrow 0$ {\it before}
the limit $\lambda \rightarrow \lambda_c$ corresponds to the limit
$x \rightarrow \infty$. In this limit one finds
$f(x) \sim {5 \over 128}~{1 \over x^2} + {\cal O}({1 \over x^3})$,
that is, the characteristic critical behavior of 2D gravity
``silently'' disappears and we recover a power series in
${1 \over x}$ corresponding, in view of (\ref{scalx}),
to a perturbative expansion in lattice defects $\beta$. In this limit
the
characteristic growth of surfaces as a function of area $A$ is
\begin{equation}
{\cal Z}(t,A,\beta)~\sim~t^4~(~A+{\cal O}(\beta^2 A^3)~).
\label{limit}
\end{equation}
The leading order corresponds precisely to the almost flat
lattices (with exactly four positive defects) studied
in our second work\cite{KSWII}.
The corrections are interpreted as insertions of negative
defects, balanced by further positive defects.
The typical shape of the surfaces in this limit is a generalisation
of the one we found for ``almost flat'' graphs:
Long, filamentary cylinders growing out from every negative
curvature defect.

(c) It is easy to prove that the scaling function (\ref{scalf}) is
the simplest possible function with the limiting properties
discussed in (a) and (b).

The above results might be interpreted in terms of
a continuum model of quantized curvature defects, in which
the localised defects move around like particles in a gas
on a flat background space.
The deficit angle, $\Delta \theta$, of a defect can take on the
values $\Delta \theta =\pi, 0$ and $-\pi$. A positive curvature
defect is
surrounded by a conical geometry, whereas a negative curvature defect
corresponds to a saddle-type insertion (see Fig.~1).
The higher order negative
curvature defects ($-2\pi$ and higher)
would not be expected to play a
role in this limit (the entropy from moving two low order defects
around would completely dominate that from a single higher order
defect).  The coupling $\beta$ can be interpreted as a fugacity
controlling the number of defects. The flat space limit
$\beta\rightarrow 0$ consists of four defects of degree $\pi$ moving
around
with respect to one another. Varying the fugacity, $\beta$, allows
one
to smoothly interpolate between flat space, (\ref{limit}) (with four
defects), and pure gravity (\ref{growth})
(with an infinite number of defects).

One might also attempt to develop this picture
directly in the continuum.
One could start with
the conformal metric of a flat surface with localised curvature
defects. It can be represented locally as
$g_{ab} = \delta_{ab}~e^{\varphi(z)}$ with
\begin{equation}
\varphi(z)=  \sum_{j=1}^M~R_j~\ln(z-z_j)^2,
\label{metr}
\end{equation}
where $R_j=-1,1$. Symbolically,
the partition function might be written as
\begin{equation}
Z(\mu,\b)=
\sum_M~\beta^M~\int d [z_1,...,z_M]
{}~e^{-\mu \int d^2 z~\sqrt{{\rm det} g(z)} } .
\label{partf}
\end{equation}
Here we introduced the fugacity of curvature defects $\beta$
instead of
the explicit $R^2$-term in the action. It serves the same purpose:
for
$\beta \rightarrow 0$
we arrive at the completely flat metric, whereas for
$\beta \sim 1$ the
system should show the behaviour of pure quantum gravity, at least in
the infrared domain. We retained the notation
$\beta$ to denote the
parameter playing a role similar to the
$R^2$ coupling in the above discrete model.

This formulation resembles a little bit the two-dimensional Coulomb
gas\index{Coulomb gas|see{Quantum gravity}}
problem. However, the measure of integration $d [z_1,...,z_M]$ of
the positions of the curvature defects is a complicated object: it
should take into account the topology of the surface and the
existence
of zero modes (the action does not depend on some directions in the
space of the $z_i$). It would be very interesting to make
this direct continuum formulation more precise.

Another interesting issue is the role of exponential corrections
appearing due to the structure of elliptic functions.
In fact all physical quantities, such as
${\cal Z}(t,\lambda,\beta)$ (see eq.(\ref{PART})),
contain exponentially small terms in the limit
$\lambda \rightarrow \lambda_c$ and $\beta \rightarrow 0$,
thus leading to an essential
singularity at $\beta=0$, $\lambda=1$. One can obtain the first
correction of this type in e.g.~the free energy
${\bar f}(\beta)$ per unit area
in the thermodynamical limit $\lambda=\lambda_c$:
\begin{equation}
{\cal Z}(t,A,\beta)~\sim~{t^4\over\beta^{9\over 2}}~
e^{{\bar f}(\beta) A}~A^{-{7 \over 2}}
\quad {\rm with} \quad
{\bar f}(\beta)=
{\pi \over \sqrt{2}} \beta \Bigl[(1+\dots )
+ e^{-{\pi \sqrt{2}\over\beta}}(4+\dots) \Bigr]
\label{expc}
\end{equation}
where
${\bar f}(\beta)=\lim_{A\rightarrow\infty}
{1\over A}\ln {\cal Z}(t,A,\beta) =\lambda_c$
and the dots denote terms of order $\beta^3$ and higher.

These exponential terms are likely to be lattice artifacts.
They emerge even in the simplest calculation for the flat
closed quadrangulation with four positive curvature defects,
where they appear as discrete corrections to the approximation of
elliptic sums by integrals close to the continuum limit.

On the other hand, formula (\ref{expc})
corresponds to the critical free energy
as a function of the curvature fugacity $\beta$
(i.e.~we have already taken
the continuum limit).
It is possible that the exponential terms might be
corrections relevant for the
statistical mechanics of random lattices at long distances
(of order
$1 \over \beta$) rather than for continuous 2D-gravity.

In conclusion, we have tried to emphasize two points
in this presentation:

1.~{\it Technical advances in large $N$ group theory
\index{Large $N$ expansion} are possible.}
Apparently hopelessly difficult problems like the
model\index{Matrix models! Hermitian in external field}
(\ref{modmatrix}) become tractable when one changes the technique.

Our approach raises a number of interesting questions.
First of all, the method we have employed seems rather indirect
and is definitely very involved. Is there a hidden
simplicity we have missed so far? Are there alternative ways
to obtain our results?
Secondly, one should ask whether the method could be adapted
to other theories describing new interesting physics; in
particular gauge theories\index{Large $N$ Expansion! and QCD}
and strings\index{Large $N$ Expansion! and string theory}
in physical dimensions.

2.~{\it Each such advance leads to the capability to address new
physics questions. The present advance
enables us to treat the problem of two-dimensional
higher curvature
gravity\index{Quantum gravity! higher curvature}.}

We have presented for the first time an exactly
solvable model interpolating between flat space and
2D quantum gravity. This addresses a long-standing open
problem. We would like to stress, however, that we have by
no means demonstrated yet the {\it universality} of our
result. Could it be that finetuning the weights $t_q$ leads to
new phases of gravity? Here we certainly do not mean to rederive
the usual multicritical phases of the one-matrix model
(\ref{matrix}). We rather envision tuning the weights in a subtle
way so as to reach a phase of {\it smooth} 2D gravity.
Certainly the analytic complexity of the result could be
a hint that much more waits to be discovered.
(Compare e.g.~the densities (\ref{density}),(\ref{moddens}) of the
model (\ref{modmatrix}) under study with the rather trivial
algebraic densities obtained in the absence of the external field.)
Before addressing this issue, we repeat, one first has
to simplify the method enough to allow for a deeper insight into the
analytic structure of the solution for general couplings
$t_q$.

\end{document}